\documentclass{PoS}

\usepackage{graphicx,amsfonts,amsmath,amssymb,amstext,bm}
\definecolor{darkred}{rgb}{0.6,0.0,0.0}
\definecolor{darkgreen}{rgb}{0.0,0.5,0.0}
\definecolor{darkblue}{rgb}{0.0,0.0,0.7}

\newcommand*\samethanks[1][\value{footnote}]{\footnotemark[#1]}

\title{The overdamped chiral magnetic wave}

\ShortTitle{The overdamped chiral magnetic wave}

\author{\speaker{I.~A.~Shovkovy}\thanks{Supported by the U.S. National Science Foundation 
under Grants No. PHY-1713950.}\\
        College of Integrative Sciences and Arts, Arizona State University, Mesa, Arizona 85212, USA\\
        Department of Physics, Arizona State University, Tempe, Arizona 85287, USA\\
        E-mail: \email{igor.shovkovy@asu.edu}}

\author{D.~O.~Rybalka\samethanks\\
        Department of Physics, Arizona State University, Tempe, Arizona 85287, USA\\
        E-mail: \email{drybalka@asu.edu}}    

\author{E.~V.~Gorbar\thanks{Supported by the Program of Fundamental Research of the Physics and Astronomy Division of the National Academy of Sciences of Ukraine.}\\
        Department of Physics, Taras Shevchenko National Kiev University, Kiev, 03680, Ukraine\\
        Bogolyubov Institute for Theoretical Physics, Kiev, 03680, Ukraine\\
        E-mail: \email{gorbar@bitp.kiev.ua}}

\abstract{About eight years ago it was predicted theoretically that a charged chiral plasma could 
support the propagation of the so-called chiral magnetic waves, which are driven by the anomalous 
chiral magnetic and chiral separation effects. This prompted intensive experimental efforts in search 
of signatures of such waves in relativistic heavy-ion collisions. In fact, several experiments have 
already reported a tentative detection of the predicted signal, albeit with a significant background 
contribution. Here, we critically reanalyze the theoretical foundations for the existence of the chiral 
magnetic waves. We find that the commonly used background-field approximation is not sufficient 
for treating the waves in hot chiral plasmas in the long-wavelength limit. Indeed, 
the back-reaction from dynamically induced electromagnetic fields turns the chiral magnetic wave 
into a diffusive mode. While the situation is slightly better in the strongly-coupled near-critical regime 
of quark-gluon plasma created in heavy-ion collisions, the chiral magnetic wave is still strongly 
overdamped due to the effects of electrical conductivity and charge diffusion.}

\FullConference{XIII Quark Confinement and the Hadron Spectrum - Confinement 2018\\
		31 July - 6 August 2018\\
		Maynooth University, Ireland}

\begin{document}

\section{Introduction}
\label{sec:intro}

Chiral relativistic plasmas can be realized in a number of physical systems at very high temperatures and/or 
densities when the masses of fermions are negligible compared to the temperature and/or chemical potential.
Typical examples include the quark-gluon plasma in heavy-ion collisions 
\cite{Liao:2014ava,Miransky:2015ava,Huang:2015oca,Kharzeev:2015znc}, the primordial plasma in the 
early Universe \cite{Rogachevskii:2017uyc,Gorbunov:2011zz}, and several types of 
degenerate forms of matter in compact stars \cite{Weber:2004kj}. The pseudo-relativistic analogs of chiral plasmas 
can be also found in Dirac and Weyl materials \cite{Vafek:2013mpa,Burkov:2015hba,Gorbar:2017lnp}. A distinctive 
feature of chiral plasmas is the presence of an approximately conserved chiral charge, which comes in addition to 
the exactly conserved electric charge (or the fermion number) and is violated only by the quantum chiral anomaly 
\cite{Adler,Bell-Jackiw}. 

The first studies of anomalous effects in chiral plasmas started several decades ago with a series of pioneering 
papers by Vilenkin \cite{Vilenkin:1978hb,Vilenkin:1979ui,Vilenkin:1980fu}. The recent revival of interest to the subject 
was triggered by the realization that such effects could be observed via the angular correlations of charged particles in relativistic 
heavy-ion experiments \cite{Kharzeev:2004ey}. It was also suggested that chiral anomalous effects could have 
profound effects on the evolution of magnetic fields in the early Universe \cite{Joyce:1997uy,Boyarsky:2011uy,
Tashiro:2012mf}. The principle anomalous processes in magnetized chiral plasmas at nonzero electric or chiral 
charge chemical potentials ($\mu$ or $\mu_5$, respectively) are the chiral separation effect (CSE) 
\cite{Vilenkin:1980fu,Metlitski:2005pr} and the chiral magnetic effect (CME) \cite{Fukushima:2008xe}.
The essence of the CSE is the induction of a nondissipative chiral current 
$\mathbf{j}^5_\textrm{CSE}=e\mathbf{B}\mu /(2\pi^2)$ when $\mu\neq 0$, and the CME is similarly characterized 
by the electric current $\mathbf{j}_\textrm{CME}=e\mathbf{B}\mu_5 /(2\pi^2 )$ when $\mu_5\neq 0$. 

About eight years ago it was proposed that the interplay of the CSE and CME in chiral plasmas can lead to the 
existence of a special type of gapless collective mode, which was called the chiral magnetic wave (CMW) 
\cite{Kharzeev:2010gd}. It was argued that the propagation of the CMW would be sustained by alternating 
oscillations of the local electric and chiral charge densities that feed into each other. Experimentally, the corresponding 
wave would manifest itself in heavy-ion collisions in the form of quadrupole correlations of charged particles
\cite{Gorbar:2011ya,Burnier:2011bf}. Moreover, over the last several years, a number of experimental 
detections of the predicted charge-dependent flow patterns have already been reported 
\cite{Ke:2012qb,Adamczyk:2013kcb,Adamczyk:2015eqo,Adam:2015vje,Sirunyan:2017tax}. 

In contrast to the original predictions, however, recently we found that the CMW should be a 
diffusive mode in the weakly coupled plasma in the long-wavelength limit \cite{Rybalka:2018uzh}. 
The novel aspect of our analysis was the rigorous treatment of dynamical electromagnetism in 
chiral plasmas. Here we review the details of the underlying physics responsible for turning the 
CMW into a diffusive mode. Also, by making use of the lattice results for transport coefficients, 
we extend the earlier analysis to the nonperturbative regime of the quark-gluon plasma in the 
range of temperatures between about $200~\mbox{MeV}$ and $350~\mbox{MeV}$. Despite a 
relatively low electrical conductivity and diffusion coefficients, our analysis shows that the CMW is an overdamped mode 
in the deconfined phase of quark-gluon plasma almost in the whole range of realistic parameters. 
In fact, the only regime with a well pronounced CMW might be realized in the case of a strongly 
coupled plasma under superstrong magnetic fields.

In the analysis below we use the units with $c=1$ and $\hbar=1$. The Minkowski metric is given by
$g_{\mu\nu}=\mbox{diag}(1,-1,-1,-1)$ and the Levi-Civita tensor $\epsilon^{\mu\nu\alpha\beta}$ is 
defined so that $\epsilon^{0123} = 1$.

\section{Hydrodynamic analysis of collective modes}
\label{sec:pre}

By definition, the CMW is a collective mode in a locally equilibrated chiral plasma. The corresponding 
dynamics is most naturally captured by using the framework of chiral hydrodynamics \cite{Son:2009tf},
which provides a coarse-grained description of a system on sufficiently large distance and time scales. 
The relevant degrees of freedom in such a regime are the conserved charges and 
their currents. In the case of a single-flavor charged chiral plasma, in particular, they include the 
energy-momentum tensor, as well as the fermion number (or electric charge) and chiral charge 
four-currents that satisfy the appropriate continuity equations, i.e., 
\begin{eqnarray}
\label{eq:cont-T}	
\partial_\nu T^{\mu\nu} &=& eF^{\mu\nu} j_\nu  , \\
\label{eq:cont-j}	
\partial_\mu j^\mu &=& 0, \\
\label{eq:cont-j5}	
\partial_\mu j_5^\mu &=& - \frac{e^2}{8\pi^2 } F^{\mu\nu} \tilde F_{\mu\nu} .
\end{eqnarray}
(Note that here we use the fermion number current $j^\mu$, which differs by a factor of $e$ from the 
electric current $j_{\rm el}^\mu \equiv e j^\mu$.) For simplicity of presentation, in this section we will 
limit our discussion to the case of a single-flavor plasma, but the generalization to a multi-flavor case is 
straightforward (see Sec.~\ref{sec:heavy-ion} below). 

By making use of the local fluid velocity $u^\mu$, the energy-momentum tensor and both
four-currents can be  decomposed into the longitudinal and transverse components as follows:
\begin{eqnarray}
\label{eq:hydro_j}
T^{\mu\nu} &=& \epsilon u^\mu u^\nu - \Delta^{\mu\nu} P + (h^\mu u^\nu + u^\mu h^\nu) + \pi^{\mu\nu},
	\\
\label{eq:hydro_j}
	j^\mu &=& n u^\mu + \nu^\mu,
	\\
\label{eq:hydro_j5}
	j_5^\mu &=& n_5 u^\mu + \nu_5^\mu,
\end{eqnarray}
where $\epsilon = T^{\mu\nu} u_\mu u_\nu$ is the energy density, 
$P = \Delta_{\mu\nu} T^{\mu\nu}/3$ is the pressure, $h^\mu = \Delta^{\mu\alpha} T^{\alpha\beta} u_\beta$ 
is the energy flow (or, equivalently, the momentum density vector). The definitions of the fermion number 
and chiral charge densities are given by $n=j^\mu u_\mu$ and $n_5 = j_5^\mu u_\mu$, respectively. 
The transverse currents, $\nu^\mu = \Delta^{\mu\nu} j_\nu$ and $\nu_5^\mu = \Delta^{\mu\nu} j_{5,\nu}$, 
are obtained by applying the projection operator $\Delta^{\mu\nu} \equiv g^{\mu\nu} - u^\mu u^\nu$. Finally, 
$\pi^{\mu\nu} = \Delta^{\mu\nu}_{\alpha\beta} T^{\alpha\beta}$ is the dissipative part of the energy-momentum 
tensor, which is defined in terms of the traceless 4-index projection operator $\Delta^{\mu\nu}_{\alpha\beta} 
= \frac{1}{2}\Delta^\mu_\alpha\Delta^\nu_\beta + \frac{1}{2}\Delta^\mu_\beta\Delta^\nu_\alpha 
- \frac{1}{3}\Delta^{\mu\nu}\Delta_{\alpha\beta}$. 

Since a collective motion in a charged plasma could provide an important
feedback via dynamically induced electromagnetic fields, the above set of hydrodynamic equations 
should be also supplemented by the Maxwell equations
\begin{equation}
\label{eq:maxwell}
	\partial_\nu F^{\nu\mu} = e j^\mu - en_{\textrm{bg}} u_{\textrm{bg}}^\mu,
\end{equation}
together with the Bianchi identity $\partial_\nu \tilde F^{\nu\mu} = 0$, where 
$\tilde{F}^{\mu\nu} = \frac{1}{2}\epsilon^{\mu\nu\alpha\beta} F_{\alpha\beta}$ is the dual field 
strength tensor. Note that Eq.~(\ref{eq:maxwell}) captures both the Gauss and Ampere laws, 
and $\rho^\mu = - en_{\textrm{bg}} u_{\textrm{bg}}^\mu$ accounts for a possible background 
of electrically charged particles.

In order to give a self-contained presentation, let us start by briefly reviewing the key details
of the CMW dynamics for a chiral plasma obtained in Ref.~\cite{Rybalka:2018uzh}. 
Since the local vorticity plays no vital role in the propagation of the CMW, we will consider only 
the case of a non-rotating plasma. In local equilibrium, such a plasma can be characterized 
by a pair of chemical potentials, $\bar\mu$ and $\bar\mu_5$, and temperature $\bar{T}$. In addition, 
when the plasma is at rest in the laboratory frame, its equilibrium hydrodynamic flow velocity is given 
by $\bar{u}^\mu = (1,0,0,0)$. 

We will assume that the background magnetic field points in the $z$ direction, i.e., $\bar{B}^\mu\equiv 
\tilde{F}^{\mu\nu} \bar{u}_\nu=(0,0,0,B)$. As is clear, there should be no electric field in 
equilibrium, i.e., $\bar{E}^\mu \equiv F^{\mu\nu} \bar{u}_\nu= 0$. Note that, in large macroscopic systems 
with a nonzero average $\bar\mu$, the absence of the electric field also implies the net electric neutrality. 
In general, the latter can be achieved by taking into account the background electric
charge of nonchiral particles, i.e., $\rho^\mu = - en_{\textrm{bg}} u_{\textrm{bg}}^\mu$, 
see Eq.~(\ref{eq:maxwell}) above. 

The propagation of collective modes through a chiral plasma is generically accompanied by the 
oscillations of all available dynamical parameters: the chemical potentials $\delta \mu$ and $\delta \mu_5$, 
the temperature $\delta T$, the flow velocity $\delta u^\mu$, as well as the electric and magnetic fields 
$\delta E^\mu$ and $\delta B^\mu$. In the linear approximation, it is justified to take them all in the form
of plain waves, i.e., $\delta X \propto e^{-ik\cdot x}$. 

As is easy to show, the time-components of all three vector quantities are nondynamical. In fact, they can 
be shown to vanish identically, i.e., $\delta u^0 = \delta B^0 = \delta E^0 = 0$, after taking into account the 
constraints $u^\mu B_\mu = u^\mu E_\mu = 0$ and $u^\mu u_\mu = 1$, as well as the explicit definition 
for the local (oscillating) electromagnetic field strength tensor in the laboratory frame, i.e., 
\begin{equation}
	F^{\mu\nu} = \epsilon^{\mu\nu\alpha\beta} \bar{u}_\alpha (\bar{B}_\beta + \delta B_\beta) 
	+ \delta E^\mu \bar{u}^\nu - \bar{u}^\mu \delta E^\nu.
\end{equation}
After using the Maxwell equation (\ref{eq:maxwell}), the linearized versions of the hydrodynamic equations
(\ref{eq:cont-T})--(\ref{eq:cont-j5}) can be rewritten in the following explicit form: 
\begin{eqnarray}
\label{eq:continuity-2}
	k_0 \delta\epsilon 
	- \frac{4}{3}\epsilon (\mathbf{k}\cdot\delta \mathbf{u}) 
	+ 2\xi_B k_0 (\mathbf{B}\cdot\delta \mathbf{u})
	- (\mathbf{k}\cdot\delta\mathbf{h})
	- ie\sigma_B (\mathbf{B}\cdot\delta \mathbf{E}) &=& 0, 
	\\
	\frac{4}{3}\epsilon k_0 \delta \mathbf{u} 
	- \frac{1}{3} \mathbf{k} \delta\epsilon 
	- \xi_B \mathbf{B} (\mathbf{k}\cdot\delta\mathbf{u}) 
	- \xi_B (\mathbf{B}\cdot \mathbf{k}) \delta \mathbf{u}
	+ k_0 \delta\mathbf{h} 
	+ i\frac{4\tau\epsilon}{15} \left(\mathbf{k}^2 \delta\mathbf{u} + \frac{1}{3}\mathbf{k}(\mathbf{k}\cdot\delta\mathbf{u}) \right) & & \nonumber
	\\
\label{eq:continuity-3}
	- ien\delta \mathbf{E}
	- ie\sigma_B (\mathbf{B}\times\delta\mathbf{B})
	- k_0 (\mathbf{B}\times\delta\mathbf{E})
	+ (\mathbf{B}\cdot\mathbf{k}) \delta\mathbf{B}
	- \mathbf{k} (\mathbf{B}\cdot\delta\mathbf{B}) &=& 0, 
	\\
	\label{eq:continuity-1}
	k_0\delta n
	- n (\mathbf{k}\cdot\delta\mathbf{u})
	- (\mathbf{k}\cdot\delta\boldsymbol{\nu}) &=& 0, \qquad
	\\
\label{eq:continuity-1.5}
	k_0\delta n_5 
	- n_5 (\mathbf{k}\cdot\delta\mathbf{u})
	- (\mathbf{k}\cdot\delta\boldsymbol{\nu}_5)
	- i\frac{e^2}{2\pi^2} (\mathbf{B}\cdot\delta\mathbf{E}) &=& 0 .
\end{eqnarray}
(Here we use the same notations as in Ref.~\cite{Rybalka:2018uzh}.)
Note that, for the visual clarity of equations, we removed the bars over the equilibrium 
quantities. We also introduced the following shorthand notations for the fluctuating parts of the momentum 
and current densities:
\begin{eqnarray}
	\delta\mathbf{h}
	&=& \frac{i\xi_\omega}{2} (\mathbf{k}\times\delta\mathbf{u})
	+ \mathbf{B} \delta\xi_B
	+ \xi_B \delta\mathbf{B},
	\\
	\delta\boldsymbol{\nu}
	&=& \frac{i\sigma_\omega}{2} (\mathbf{k}\times\delta\mathbf{u})
	+ \delta\sigma_B \mathbf{B}
	+ \sigma_B \delta\mathbf{B} 
	- i\frac{\tau}{3} \mathbf{k} \delta n 
	+ i\tau n k_0\delta\mathbf{u}
	+ \frac{1}{e} \sigma_E \left[\delta\mathbf{E} + (\delta\mathbf{u}\times\mathbf{B}) \right],
	\\
	\delta\boldsymbol{\nu}_5
	&=& \frac{i\sigma_\omega^5}{2} (\mathbf{k}\times\delta\mathbf{u})
	+ \delta\sigma_B^5 \mathbf{B}
	+ \sigma_B^5 \delta\mathbf{B}
	- i\frac{\tau}{3} \mathbf{k} \delta n_5
	+ i\tau n_5 k_0\delta\mathbf{u}
	+ \frac{1}{e} \sigma_E^5 \left[\delta\mathbf{E} + (\delta\mathbf{u}\times\mathbf{B}) \right].
\end{eqnarray}
For completeness, let us note that the linearized Maxwell equations take the form
\begin{eqnarray}
\label{eq:continuity-4}
	(\mathbf{k}\cdot\delta\mathbf{E}) 
	+ ie\delta n 
	+ ie\sigma_B(\mathbf{B}\cdot\delta\mathbf{u}) &=& 0,
	\\
\label{eq:continuity-5}
	(\mathbf{k}\times\delta\mathbf{B})
	+ k_0 \delta\mathbf{E}
	+ ien\delta\mathbf{u}
	+ ie\delta\boldsymbol{\nu} &=& 0,
	\\
\label{eq:continuity-6}
	- (\mathbf{k}\times\delta\mathbf{E})
	+ k_0\delta\mathbf{B} &=& 0,
	\\
\label{eq:continuity-7}
	(\mathbf{k}\cdot\delta\mathbf{B}) &=& 0 .
\end{eqnarray}
As we see, after taking the Faraday's law (\ref{eq:continuity-6}) into account, the dynamical 
oscillations of the magnetic field $\delta\mathbf{B}$ can be expressed in terms of the electric 
field $\delta\mathbf{E}$, namely $\delta\mathbf{B}= (\mathbf{k} \times \delta\mathbf{E})/k_0$. 
In such a form, the latter also automatically satisfies Eq.~(\ref{eq:continuity-7}), provided $k_0\neq 0$.

For simplicity, in this section we assume that all dissipative processes in the system are controlled by the 
same phenomenological relaxation-time parameter $\tau$. In the case of electrical conductivity, for example,
one can use $\sigma_E\simeq e^2 \tau \chi/3$, where $\chi=\partial n/\partial\mu$ is the fermion number 
susceptibility \cite{Gorbar:2016qfh}. Similarly, the chiral counterpart of conductivity $\sigma_E^5$ 
can be given as $\sigma_E^5\simeq e^2 \tau \chi^\prime$, where $\chi^\prime=\partial n_5/\partial\mu$
\cite{Gorbar:2016qfh}. 

In connection to the CMW dynamics, the most important transport coefficients are $\sigma_B=e \mu_5/(2\pi^2)$ 
and $\sigma_B^5=e \mu/(2\pi^2)$, which originate from the chiral anomaly and are responsible for the 
CME and CSE, respectively. For the definition of other transport coefficients (i.e., $\xi_B$, $\xi_\omega$, 
$\sigma_\omega$, and $\sigma^5_\omega$) and their role in the dynamics of collective modes, see 
Ref.~\cite{Rybalka:2018uzh}. 

The complete analysis of the linearized system of equations is quite involved in general and will not be 
repeated here. An interested reader is referred to the detailed study in Ref.~\cite{Rybalka:2018uzh}. Here 
it will suffice to mention that the spectrum of {\em propagating} modes contains only the sound and Alfv\'en 
waves in the regime of high temperature, and the plasmons and helicons at high density. All other 
modes, including the CMW are strongly overdamped, or completely diffusive. In the rest of these proceedings, 
we will concentrate our attention exclusively on the chiral magnetic wave and discuss the underlying reasons
for its overdamped nature.

\section{Chiral magnetic wave in high temperature plasma}
\label{sec:high_T}

Since one of the most interesting applications of the CMW was proposed in the context of relativistic 
heavy-ion collisions, it is instructive to start our analysis from the case of chiral plasma in the regime 
of high temperature. In order to sort out the key details of underlying physics, however, it will be 
illuminating to first consider the simplest case of a chiral plasma made of single flavor massless 
fermions. It will be also instructive to start from the case of a weakly interacting case (which is 
realized, for example, at sufficiently high temperatures). As we will see in Sec.~\ref{sec:heavy-ion}, 
the key details of the analysis are similar also in the nonperturbative regime of the strongly-interacting 
quark-gluon plasma with several light flavors.

In order to model the conditions in the plasma produced by relativistic heavy-ion collisions, where the 
typical values of the chemical potentials are much smaller than the temperature, it is sufficient to set 
$\mu=\mu_5=0$ in our analysis. (For the quantitative effects of a small nonzero chemical potential $\mu$ 
in the high-temperature regime, see Ref.~\cite{Rybalka:2018uzh}.) In the case of the vanishing chemical 
potentials, the system of hydrodynamic equations takes the following simpler form:
\begin{eqnarray} 
\label{eq-energy-X}
	k_0 \delta\epsilon 
	- \frac{4}{3}\epsilon (\mathbf{k}\cdot\delta \mathbf{u}) = 0,&&
	\\
\label{eq-momentum-X}
	\frac{4}{3}\epsilon k_0 \delta \mathbf{u}
	- \frac{1}{3} \mathbf{k} \delta\epsilon 
	+ i\frac{4\tau\epsilon}{15} \left(\mathbf{k}^2 \delta \mathbf{u} + \frac{1}{3}\mathbf{k}(\mathbf{k}\cdot\delta \mathbf{u})\right) 
	- k_0 (\mathbf{B}\times\delta \mathbf{E})
	+ (\mathbf{B}\cdot \mathbf{k}) \delta \mathbf{B}
	 	- \mathbf{k} (\mathbf{B}\cdot\delta \mathbf{B}) = 0,&&
		\\
\label{eq-electric-X}
	k_0\delta n
	- (\mathbf{B}\cdot \mathbf{k}) \delta\sigma_B 
	+ i\frac{\tau}{3} \mathbf{k}^2 \delta n
	- \frac{1}{e} \sigma_E (\mathbf{k}\cdot\delta \mathbf{E}) 
	- \frac{1}{e} \sigma_E (\mathbf{k}\cdot(\delta \mathbf{u}\times\mathbf{B})) = 0,&\qquad&
	\\
\label{eq-chiral-X}
	k_0\delta n_5
	- (\mathbf{B}\cdot \mathbf{k}) \delta\sigma^5_B 
	+ i\frac{\tau}{3} \mathbf{k}^2 \delta n_5
	- i\frac{e^2}{2\pi^2 } (\mathbf{B}\cdot\delta \mathbf{E}) = 0.&&
\end{eqnarray}
The corresponding Maxwell equations read
\begin{eqnarray} 
	(\mathbf{k}\cdot\delta \mathbf{E}) 
	+ ie\delta n &=& 0,
	\\
	\frac{\mathbf{k}}{k_0}(\mathbf{k}\cdot\delta \mathbf{E}) 
	+ \frac{1}{k_0}(k_0^2-\mathbf{k}^2) \delta \mathbf{E}
	+ ie\mathbf{B} \delta\sigma_B 
	+ e\frac{\tau}{3} \mathbf{k} \delta n  
	+ i\sigma_E \left[\delta \mathbf{E} + (\delta \mathbf{u}\times\mathbf{B}) \right] &=& 0 , 
\end{eqnarray}
where we took into account that $\delta\mathbf{B}= (\mathbf{k} \times \delta\mathbf{E})/k_0$.

After carefully examining the general structure of the above coupled set of equations, we find that the
system can be factorized into two blocks. In particular, the independent variables in one of the blocks
can be chosen as follows: $\delta\mu$, $\delta\mu_5$, $(\mathbf{k}\cdot\delta\mathbf{E})$, 
$(\mathbf{B}\cdot\delta\mathbf{E})$, and $((\mathbf{k}\times\mathbf{B})\cdot\delta\mathbf{u})$. 
This is the block that describes the would-be CMW among other eigenmodes. 

Because of the specific dependence of the CSE and CME currents on the magnetic field, i.e., 
$\mathbf{j}^5_\textrm{CSE}=\mu e\mathbf{B}/(2\pi^2)$ and 
$\mathbf{j}_\textrm{CME}=\mu_5 e\mathbf{B}/(2\pi^2 )$, 
the propagation of the CMW is most prominent in the direction of the magnetic field. This is also 
clear from Eqs.~(\ref{eq-electric-X}) and (\ref{eq-chiral-X}), where the CSE and CME are captured 
by the terms proportional to $(\mathbf{B}\cdot \mathbf{k})$. For the purposes of our study, therefore, 
it is sufficient to concentrate only on the case with the wave vector $\mathbf{k}$ parallel to the 
background magnetic field $\mathbf{B}$. Then, the equations for the CMW greatly simplify, i.e., 
\begin{eqnarray}
\label{eq:sys-1}
	k_0\delta n
	- k B \delta\sigma_B 
	+ i\frac{\tau}{3} k^2 \delta n
	- \frac{1}{e} \sigma_E k \delta E_z
	&=& 0,
	\\
\label{eq:sys-2}
	k_0\delta n_5
	- k B \delta\sigma^5_B
	+ i\frac{\tau}{3} k^2 \delta n_5
	- i\frac{e^2}{2\pi^2 } B \delta E_z &=& 0,
	\\
\label{eq:sys-3}
	k \delta E_z
	+ ie\delta n &=& 0.
\end{eqnarray}
It might be instructive to emphasize that these equations do not contain any dependence on the oscillations 
of the fluid velocity $\delta \mathbf{u}$. This is the consequence of assuming $\mathbf{k}\parallel \mathbf{B}$
and is not true in general for the CMW with an arbitrary direction of propagation.

After taking into account the explicit expressions for the number density and chiral charge density 
susceptibilities, $\chi=\partial n/\partial\mu$ and $\chi_5=\partial n_5/\partial\mu_5$, in the high-temperature 
plasma we find
\begin{eqnarray}
	\frac{\delta\sigma_B}{\delta n_5} &=& \frac{e}{2\pi^2\chi_5}= \frac{3e}{2\pi^2T^2}, \\
	 \frac{\delta\sigma^5_B}{\delta n} &=& \frac{e}{2\pi^2\chi}= \frac{3e}{2\pi^2T^2}.
\end{eqnarray}
By making use of these relations, and eliminating the electric field $\delta E_z$ with the help of Gauss's law 
(\ref{eq:sys-3}), we then derive the following system of equations: 
\begin{eqnarray}
	\left(k_0 + i\frac{\tau}{3} k^2 + i\sigma_E \right) \delta n
	- \frac{3eB k}{2\pi^2T^2} \delta n_5
	&=& 0,
	\\
	- \left(\frac{3eB k}{2\pi^2T^2} + \frac{e^3B}{2\pi^2 k} \right)\delta n 
	+ \left(k_0 + i\frac{\tau}{3} k^2 \right)\delta n_5 &=& 0. 
\end{eqnarray}
By solving the corresponding characteristic equation, we finally obtain the spectrum of collective modes 
\begin{equation}
\label{eq:dispersion}
	k_0^{(\pm)} = - i \frac{\sigma_E}{2}
	\pm i \frac{\sigma_E}{2} \sqrt{1- \left(\frac{3eB}{\pi^2T^2\sigma_E}\right)^2 
	\left(k^2+ \frac{e^2 T^2}{3} \right)} - i\frac{\tau}{3} k^2.
\end{equation}
It might be appropriate to mention here that a similar dependence of the CMW energy on the 
electrical conductivity was also obtained in Ref.~\cite{Abbasi:2016rds}. As is clear, the collective 
modes are diffusive when the expression under the square root is positive, i.e., when the following 
condition is satisfied:
\begin{equation}
\frac{eB}{\pi^2\sigma_E\sqrt{\chi\chi_5}} \sqrt{k^2+ e^2 \chi} 
=\frac{3eB}{\pi^2T^2\sigma_E} \sqrt{k^2+ \frac{e^2 T^2}{3}} <1 .
\label{condition-diffusive}
\end{equation}
For the long wavelength modes with $k\lesssim eT$, this inequality is easily satisfied in 
sufficiently hot plasmas and/or for sufficiently weak background magnetic fields. 

In fact, in the case of weakly coupled plasmas, Eq.~(\ref{condition-diffusive}) always holds true 
when the hydrodynamic limit is realized. Indeed, at weak coupling, the validity of hydrodynamics 
implies the following hierarchy of scales: $l_d \ll l_B\lesssim l_{\rm mfp} \ll \lambda_k$, 
where $l_{d} \simeq 1/T$ is de Broglie wavelength, $l_B=1/\sqrt{eB}$ is the magnetic length, 
$l_{\rm mfp} \simeq \tau \sim l_{d}/e^2$ is the particle mean free path, and $\lambda_k\simeq 2\pi/k$ 
is the characteristic wavelength of the hydrodynamic modes. Note also that the electrical conductivity 
scales as $\sigma_E \sim T/(e^2\ln e^{-1})$ at weak coupling  \cite{Arnold:2000dr}. 

The situation in the near-critical regime of the quark-gluon plasma created in the relativistic heavy-ion 
collisions is not as simple, however. First of all, because of strong coupling, there is no clear separation 
between the relevant length scales, $l_d \simeq l_{\rm mfp}$. Additional complications arise from the 
fact that the plasma is created in a rather small region of space. Nevertheless, the hydrodynamic 
description is expected to be suitable for such finite-size fireballs of quark-gluon plasma. The 
quantitative analysis of the corresponding case will be presented in Sec.~\ref{sec:heavy-ion}.

It is instructive to study the physical reasons for the diffusive nature of the collective modes in
Eq.~(\ref{eq:dispersion}) in the case of chiral plasmas at sufficiently high temperature and/or sufficiently 
weak background magnetic fields, i.e., when the expression on the left-hand side of Eq.~(\ref{condition-diffusive}) 
is much smaller than $1$. Out of the two modes in Eq.~(\ref{eq:dispersion}), the first one 
has a smaller imaginary part, i.e., 
\begin{equation}
k_0^{(+)} \simeq - \frac{i}{\sigma_E} \left(\frac{3eB}{2\pi^2T^2}\right)^2 
	\left(k^2+ \frac{e^2 T^2}{3} \right) - i\frac{\tau}{3} k^2 ,
\end{equation}
and describes the chiral charge diffusion, with a small admixture of an induced 
electric charge,
\begin{equation}
\delta n^{(+)} \simeq - i \frac{3eB}{2\pi^2T^2\sigma_E}k\delta n^{(+)}_5 .
\end{equation}
The other mode has a larger imaginary part, which is determined almost completely by the electrical 
conductivity, i.e., 
\begin{equation}
k_0^{(-)} \simeq - i \sigma_E + \frac{i}{\sigma_E} \left(\frac{3eB}{2\pi^2T^2}\right)^2 
	\left(k^2+ \frac{e^2 T^2}{3} \right) - i\frac{\tau}{3} k^2 ,
\end{equation}
and describes the electric charge diffusion, with a small admixture of an induced 
chiral charge, 
\begin{equation}
\delta n^{(-)}_5 \simeq i \frac{3eB}{2\pi^2T^2\sigma_E}  \left(k+\frac{e^2 T^2}{3k}\right) \delta n^{(-)} .
\end{equation}
Clearly, neither of the two modes resembles the conventional CMW with the expected dispersion 
relation $k^{\rm (CMW)}_0 = \pm v_{\rm CMW} k$, where $v_{\rm CMW} =3eB/(2\pi^2T^2)$ 
obtained in the background-field approximation \cite{Kharzeev:2010gd}. As discussed in detail 
in Ref.~\cite{Rybalka:2018uzh}, the dramatic difference is the result of carefully taking dynamical 
electromagnetism into account. In fact, it is the high electrical conductivity of the plasma that plays 
the most important role. This can be explicitly verified by considering the limit $\sigma_E\to 0$ in 
Eqs.~(\ref{eq:sys-1}) and (\ref{eq:sys-2}). In such a formal limit, the dispersion relations become
\begin{equation}
\label{eq:dispersion_collisionless}
	k_0^{(\pm)} = \pm \frac{3eB}{2\pi^2T^2} \sqrt{k^2  + \frac{e^2 T^2}{3 }}- i\frac{\tau}{3} k^2 ,
	\quad \mbox{when} \quad \sigma_E\to 0.
\end{equation}
These describe a pair of propagating CMW modes, although they are not the conventional ones
because of a nonzero energy gap in the spectrum. The origin of the gap can be traced to the 
last term on the left-hand side of Eq.~(\ref{eq:sys-2}), which is the usual chiral anomaly term 
proportional to $B \delta E_z$. It gives a nontrivial contribution after the Gauss law (\ref{eq:sys-3}) 
is taken into account. So, strictly speaking, the gap is the result of dynamical electromagnetism 
as well. Note that the gap in the energy spectrum of the CMW was also found in the context of 
Weyl semimetals in Ref.~\cite{Gorbar:2018nmg}.

From a physics viewpoint, the detrimental role of electrical conductivity on the propagation of the 
CMW can be relatively easily understood. The fundamental time scale for the CMW is set by the 
CSE and CME, which convert the oscillating electric and chiral charge densities into each other. 
The corresponding time is $t_\textrm{CMW}\simeq 2\pi^2T^2/(3eBk)$. However, at sufficiently 
high temperatures and/or low magnetic fields, this is much longer 
than the time scale for screening of the electric charge fluctuations due to the electrical conductivity, 
$t_\textrm{scr} \simeq\sigma_E^{-1}$. As a result, any local charge perturbation dissipates much 
quicker than the time it takes to produce a substantial chiral charge imbalance to sustain the CMW. 

\section{Chiral magnetic wave in heavy-ion collisions}
\label{sec:heavy-ion}

As we mentioned in the previous section, in the case of strongly coupled quark-gluon plasma 
created in the relativistic heavy-ion collisions, the analysis is not so simple because there is no 
clear separation between the relevant length scales in the problem. One also has to take into
account the effects associated with a small size of the system, its finite life-time, and to use  
realistic values for the transport coefficients. Here we perform the corresponding study in the 
nonperturbative regime of the quark-gluon plasma by using the transport coefficients obtained 
in lattice calculations \cite{Aarts:2007wj,Amato:2013naa,Aarts:2014nba}. 

Let us start by writing down the complete set of chiral hydrodynamic equations for the plasma 
made of two light quark flavors,
\begin{eqnarray}	
\label{eq-n-2flavor}
\partial_\mu j_f^\mu &=& 0,  
	\\
\label{eq-n5-2flavor}
\partial_\mu j_{f,5}^\mu &=& - \frac{e^2q_f^2}{8\pi^2 } F^{\mu\nu} \tilde F_{\mu\nu} , 
	\\
\label{eq-T-2flavor}
\partial_\nu T^{\mu\nu} &=& e F^{\mu\nu}  \sum_f q_f j_{f,\nu}  ,
\end{eqnarray}
where $f=u,d$, and the quark charges are $q_u= 2/3$ and $q_d= -1/3$. Note that the total electric 
current is given in terms of the individual flavor number density currents as follows:
$ j_{\rm el}^{\mu} = e \sum_f q_f  j_{f}^{\mu} $. For simplicity, we ignore the effects of the strange quark, 
which is considerably more massive than the two light quarks. It can be checked, however, that the 
results do not change much even if the strange quarks are included either as (i) an additional massless flavor 
that contributes to both sets of continuity relations (\ref{eq-n-2flavor}) and (\ref{eq-n5-2flavor}), or (ii) 
as a sufficiently massive flavor that contributes to Eq.~(\ref{eq-n-2flavor}), but not to Eq.~(\ref{eq-n5-2flavor}).

The complete set of linearized equations that describes the longitudinal CMW in the multi-flavor 
quark-gluon plasma reads
\begin{eqnarray}
\label{eq:sys-HIC-1}
	k_0\delta n_f
	-  \frac{e q_f B k}{2\pi^2 \chi_{f,5}} \delta n_{f,5} 
	+ i D_f k^2 \delta n_f
	- \frac{1}{e q_f} \sigma_{E,f} k \delta E_z
	&=& 0,
	\\
\label{eq:sys-HIC-2}
	k_0\delta n_{f,5}
	- \frac{e q_f B k}{2\pi^2  \chi_f} \delta n_{f} 
	+ i D_f k^2 \delta n_{f,5}
	- i\frac{e^2q_f^2}{2\pi^2 } B \delta E_z &=& 0,
	\\
\label{eq:sys-HIC-3}
	k \delta E_z
	+ ie \sum_f q_f \delta  n_f &=& 0,
\end{eqnarray}
where we used the following relations:
\begin{eqnarray}
\delta\sigma_{B,f} &\equiv & \frac{e q_f}{2\pi^2 } \delta \mu_{f,5} = \frac{e q_f}{2\pi^2 \chi_{f,5}} \delta n_{f,5} ,
	\\
\delta\sigma^5_{B,f}  &\equiv & \frac{e q_f}{2\pi^2 } \delta \mu_{f} =\frac{e q_f}{2\pi^2  \chi_f} \delta n_{f} ,
\end{eqnarray}
which are given in terms of the fermion number and chiral charge susceptibilities 
$\chi_f \equiv \partial n_{f}/\partial \mu_{f}$ and $\chi_{f,5}\equiv \partial n_{f,5}/\partial \mu_{f,5}$.

In the continuity relations for the flavor number charge, we also used the partial flavor contributions 
to the electrical conductivity, i.e., $\sigma_{E,f} = c_\sigma e^2 q_f^2 T$. Note that the total conductivity 
$ \sigma_{E}$ takes the form
\begin{equation}
\label{eq-conductivity-lattice}
\sigma_{E} =  \sum_f \sigma_{E,f}  = c_\sigma C^{\ell}_{\rm em} T,
\end{equation}
where $C^{\ell}_{\rm em}  = e^2 \sum_f q_f^2
= 5e^2 /9\approx  5.1\times 10^{-2}$, where we took into account the definition of the fine structure constant, 
$e^2/(4\pi)=1/137$. In the case of deconfined quark-gluon plasma, the numerical coefficient $c_\sigma$ was 
obtained in lattice calculations \cite{Aarts:2007wj,Amato:2013naa,Aarts:2014nba}. According to the most recent 
calculation \cite{Aarts:2014nba}, its value ranges from about $c_\sigma\approx 0.111$ at $T=200~\mbox{MeV}$ 
to about $c_\sigma\approx 0.316$ at $T=350~\mbox{MeV}$, see Table~\ref{table:1}. In the study of collective 
modes below, we will use these lattice values for the transport coefficients. 

\begin{table}[h!]
\centering
\begin{tabular}{ c | c | c | c}
 $T$ & $c_\sigma$ & $c_\chi$ & $c_D$ \\ 
\hline
 200~MeV & 0.111 & 0.804 & 0.758 \\ 
 235~MeV  & 0.214 & 0.885 & 1.394 \\  
 350~MeV  & 0.316 & 0.871  & 1.826   
\end{tabular}
\caption{Numerical values of coefficients $c_\sigma$, $c_\chi$, and $c_D$ at three fixed 
temperatures obtained from lattice calculations in Ref.~\cite{Aarts:2014nba}.}
\label{table:1}
\end{table}

We will also use the lattice results for the light-flavor number density susceptibilities $\chi_{f} $ and the 
diffusion coefficients $D_{f} $ \cite{Aarts:2007wj,Amato:2013naa,Aarts:2014nba}, i.e.,
\begin{eqnarray}
\label{eq-susceptibilities-lattice}
\chi_{f} &=&  c_\chi \chi^{(SB)}_{f}  ,\\
\label{eq-diffusion-lattice}
D_{f} &=& \frac{c_D  }{2\pi T},
\end{eqnarray}
where the values of numerical coefficients are flavor independent (for the light $u$- and $d$-quarks) 
and are given in Table~\ref{table:1}. Note that the Stefan-Boltzmann expression for the susceptibility 
is $\chi^{(SB)}_{f} \equiv T^2/3$. We will assume that the chiral charge susceptibility is the same as 
the fermion number one, i.e., $\chi_{f,5} = \chi_{f}$. 

While the structure of Eqs.~(\ref{eq:sys-HIC-1})--(\ref{eq:sys-HIC-3}) is very similar to 
Eqs.~(\ref{eq:sys-1})--(\ref{eq:sys-3}), one should note that the total number of coupled 
equations is larger because the fermion number and chiral charges for each flavor
satisfy independent continuity relations. With a larger number of equations, unfortunately, 
the characteristic equation becomes more complicated and no simple analytical solutions 
can be presented. Nevertheless, by making use of the intuition gained in the simpler model 
in Sec.~\ref{sec:high_T}, it is straightforward to check numerically that the underlying physics 
remains essentially the same. 

In application to quark-gluon plasma created in heavy-ion collisions, it is important to take into
account a relatively small size of the system. Such a size plays an important role as it sets an 
upper bound for the wavelengths of collective modes that could be realized, i.e., $\lambda_k
\lesssim R$, where $R$ is the system size. This implies, in turn, that there is an unavoidable 
lower bound for the values of wave vectors, $k\gtrsim 2\pi/R$. In the numerical analysis below, 
we will assume that the size of the system lies between about $12~\mbox{fm}$ and $24~\mbox{fm}$.
This would translate into an infrared cutoff for the possible wave vectors of about $100~\mbox{MeV}$ 
at $R\simeq 12~\mbox{fm}$ and $50~\mbox{MeV}$ at $R\simeq 24~\mbox{fm}$. 

Of course, there is also an upper bound for the values of wave vectors of collective modes. It is set 
by the inverse mean free part of the system. In the case of the deconfined quark-gluon plasma in
the near-critical region, the latter is likely to be of the order of $1~\mbox{fm}$ or so. For our purposes,
however, it will be sufficient to consider the wavelength $\lambda_k \gtrsim 2~\mbox{fm}$, which 
translates into the upper limit for the wave vectors $k \lesssim  600~\mbox{MeV}$. 

\begin{figure}[t]
    \includegraphics[width=0.47\textwidth]{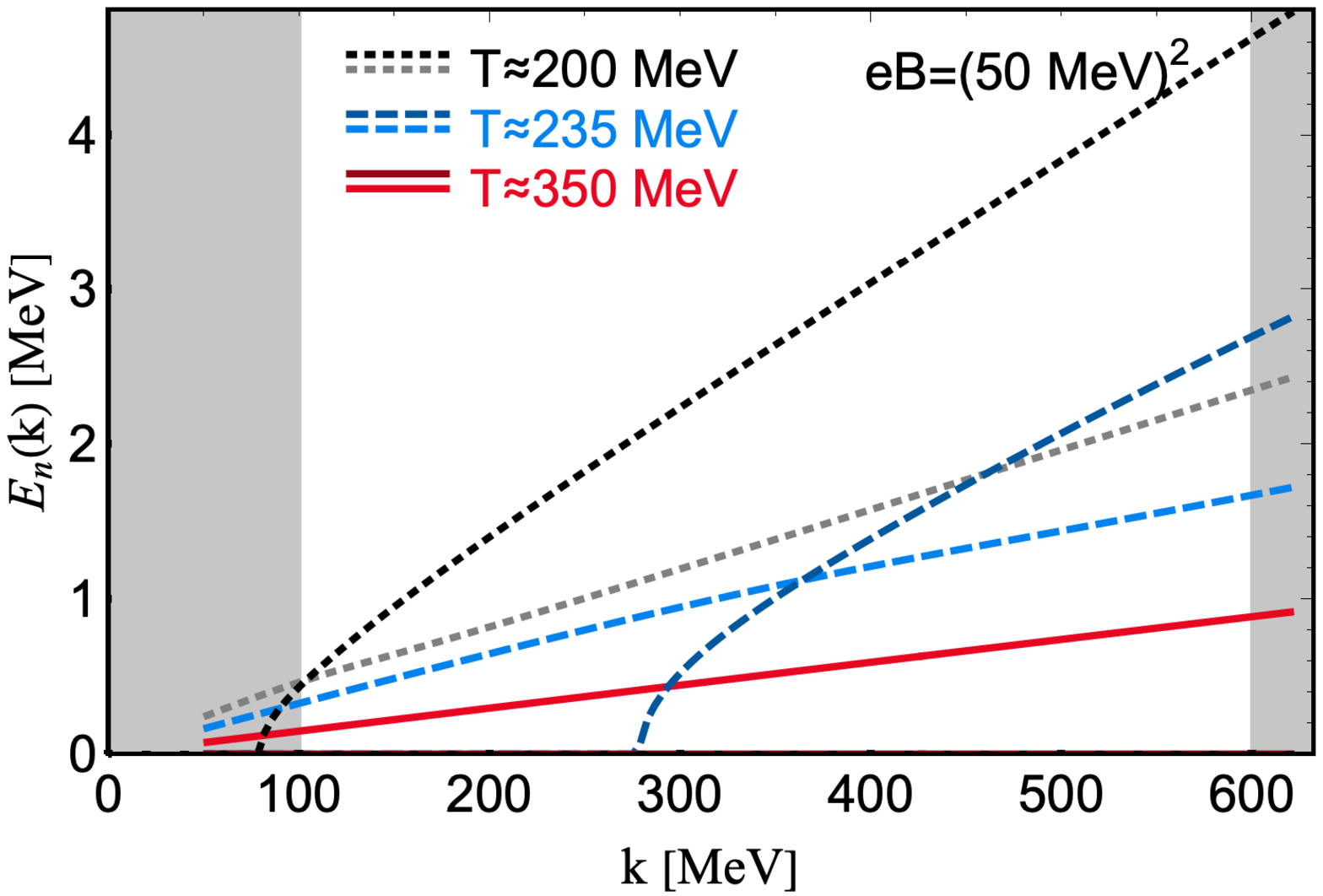}\hspace{0.02\textwidth}
    \includegraphics[width=0.49\textwidth]{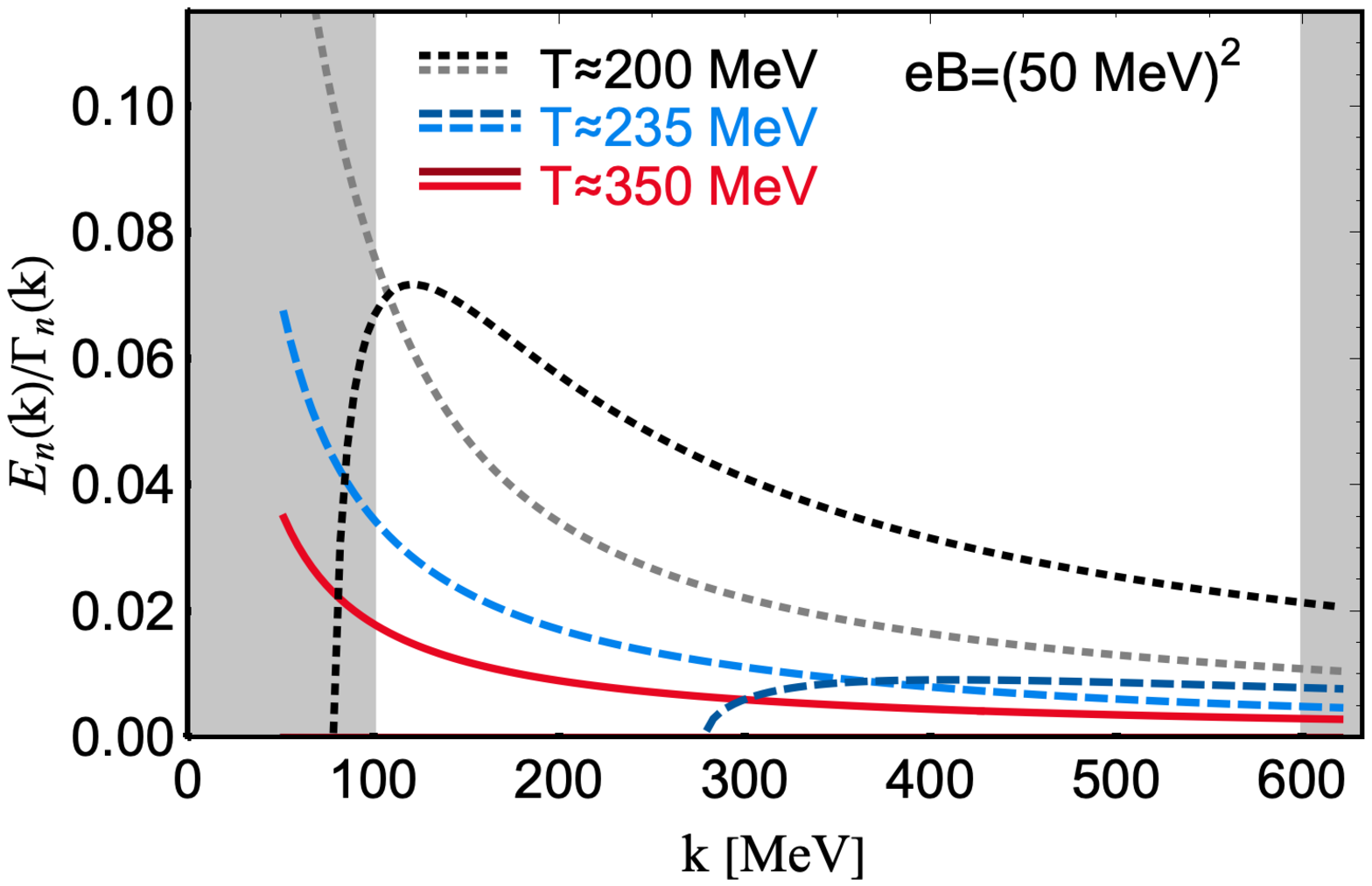}\\
    \includegraphics[width=0.48\textwidth]{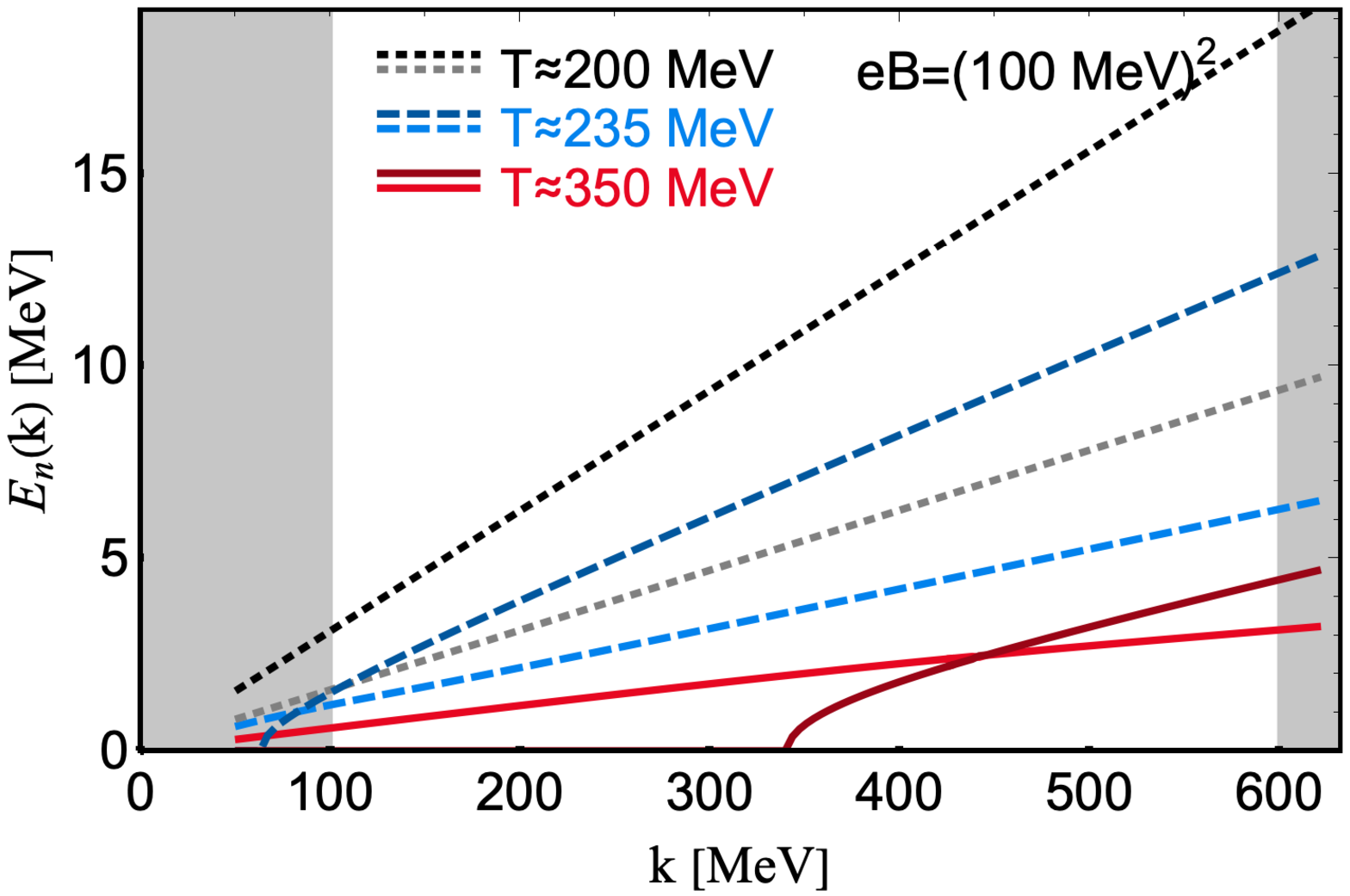}\hspace{0.02\textwidth}
    \includegraphics[width=0.48\textwidth]{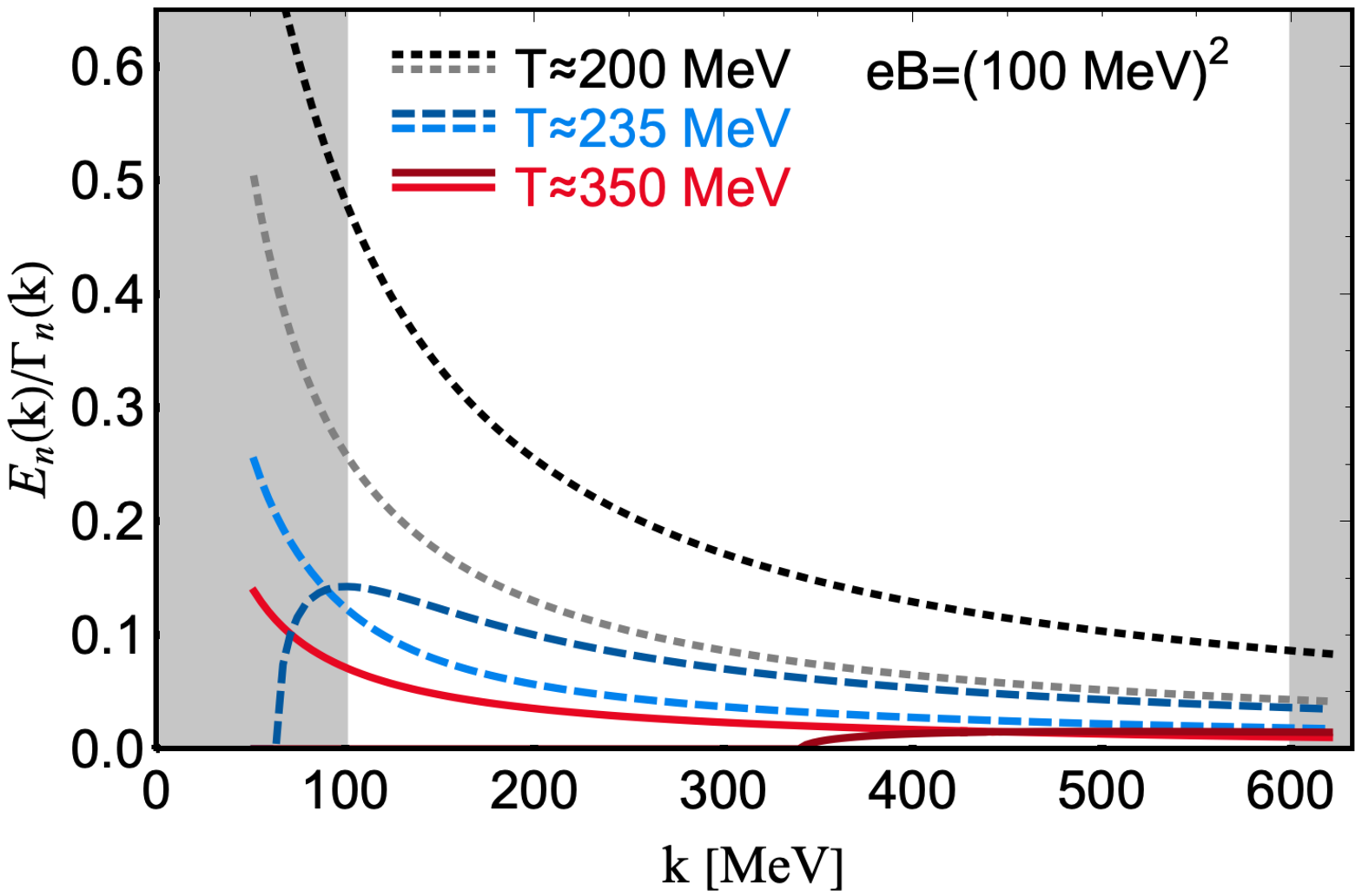}\\
    \includegraphics[width=0.48\textwidth]{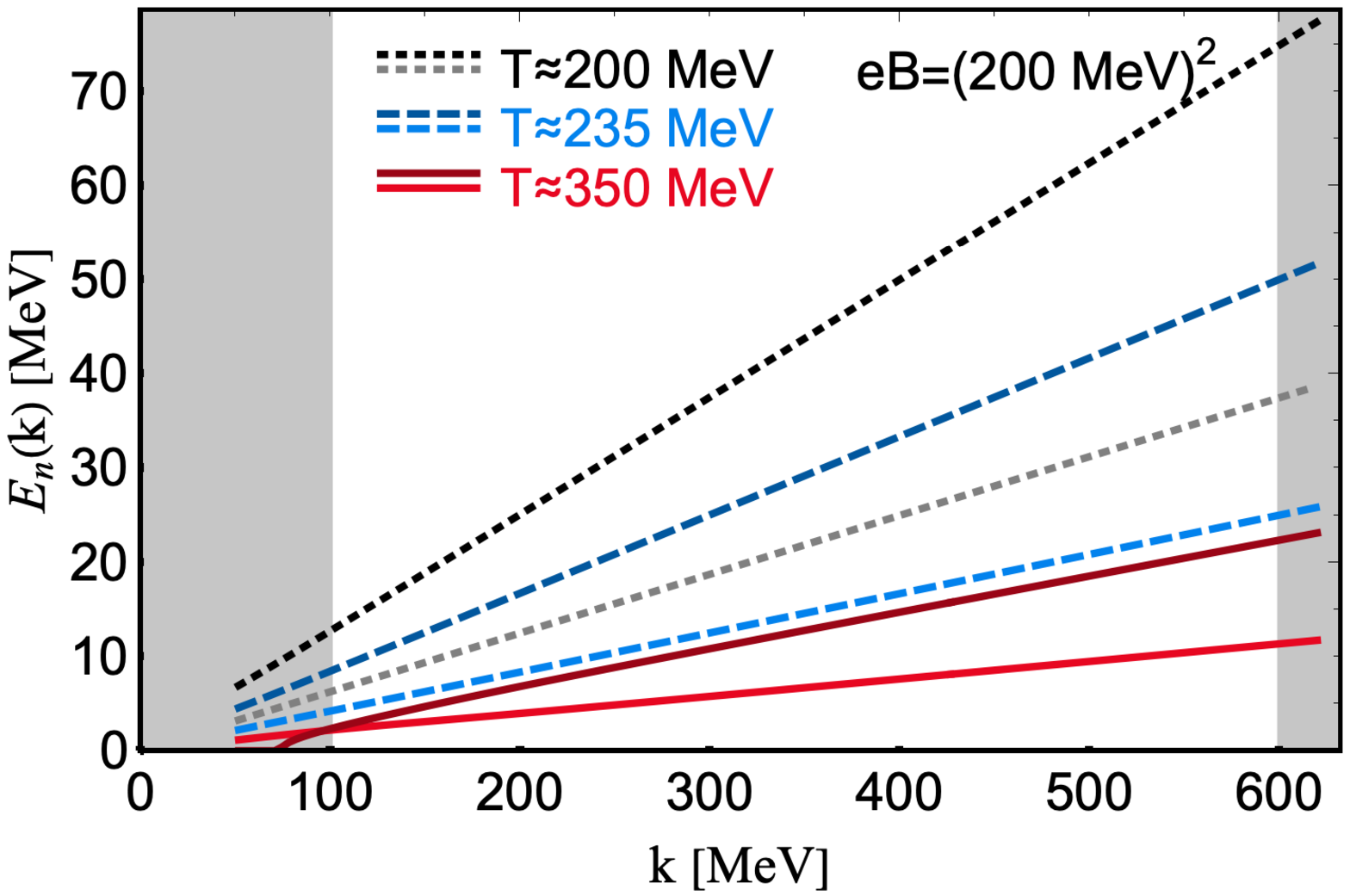}\hspace{0.02\textwidth}
    \includegraphics[width=0.48\textwidth]{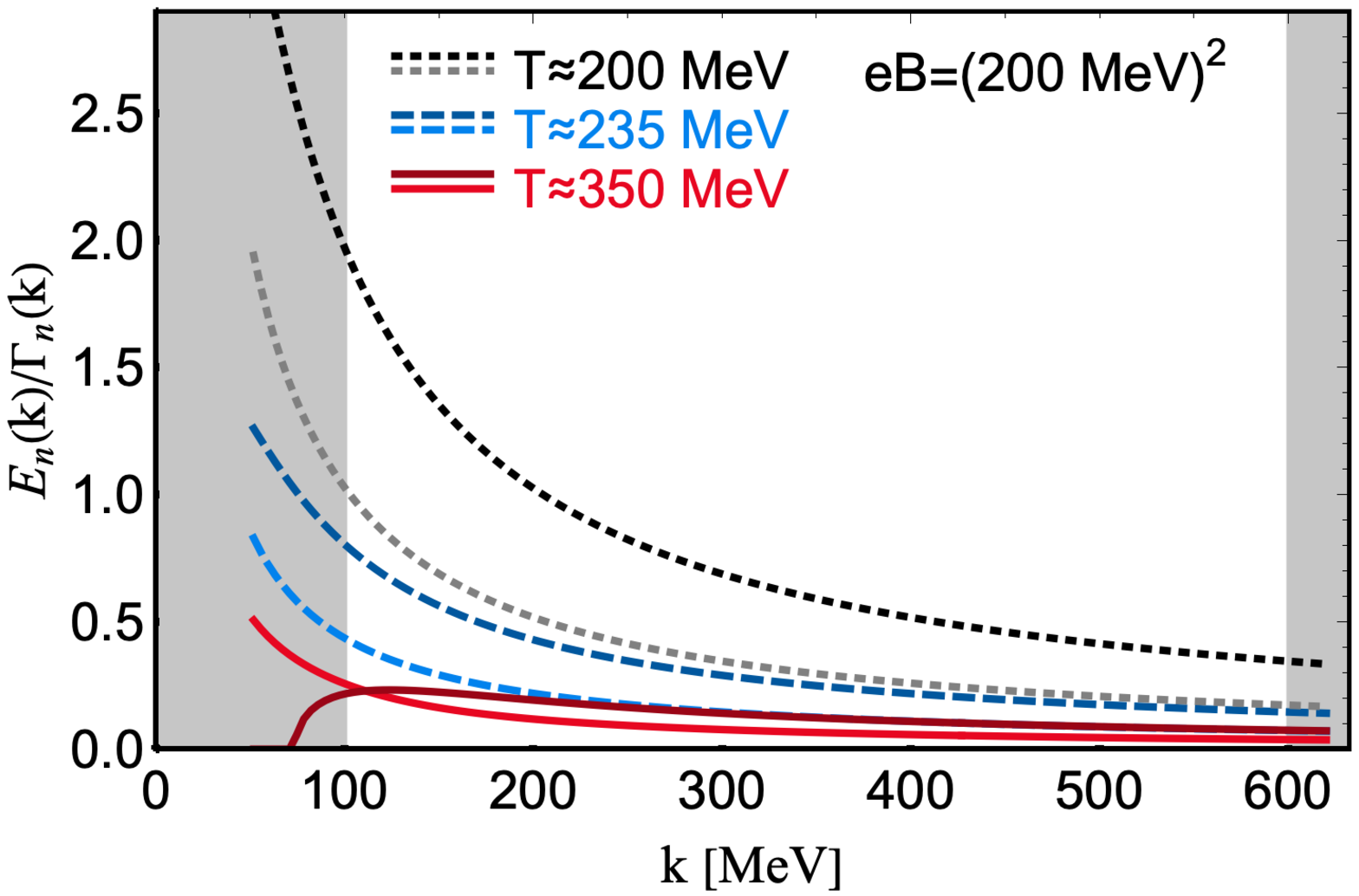}    
    \caption{The real parts of the energies (left panels) and the ratios of the real to imaginary parts of the 
    energies (right panels) of the CMW-type collective modes at three fixed values of temperature. The 
    three rows of panels show the results for three choices of the magnetic field, i.e., $ eB = (50~\mbox{MeV})^2$,
    $(100~\mbox{MeV})^2$, and $(200~\mbox{MeV})^2$, respectively. In the gray shaded regions, the 
    wavelengths lie outside the range $2~\mbox{fm}\lesssim \lambda_k \lesssim  12~\mbox{fm}$.
    The actual results are plotted down to the wave vectors as small as $k \approx 50~\mbox{MeV}$, 
    which corresponds to $\lambda_k \lesssim  24~\mbox{fm}$.}
\label{fig:modes}
\end{figure}

The numerical analysis reveals that there are two pairs of overdamped collective modes. 
The dispersion relations for both modes take the following general form:
\begin{equation}
 k_{0,n}^{(\pm)} = \pm E_{n}(k) - i   \Gamma_{n}(k), \quad \mbox{with} \quad n=1,2,
\end{equation}
where $E_{n}(k)$ and $\Gamma_{n}(k)$ are real and imaginary parts of the energies of collective modes. 
It is interesting to note that, in the long wavelength regime, one of the modes is the usual CMW, while the 
other corresponds to electrically neutral oscillations with $n_d\approx 2n_u$. 
The numerical results for the corresponding dispersion relations are summarized in Fig.~\ref{fig:modes},
where we show the dependence of the real parts of the energies, as well as the ratios of the real to 
imaginary parts, on the wave vector $k$ for three fixed values of temperature $T = 200~\mbox{MeV}$, 
$235~\mbox{MeV}$, and $350~\mbox{MeV}$, and for three fixed values of the background magnetic field, i.e., 
$eB = (50~\mbox{MeV})^2$, $(100~\mbox{MeV})^2$, and $(200~\mbox{MeV})^2$. The numerical data 
is presented for the wave vectors in the range $50~\mbox{MeV} \lesssim  k \lesssim  620~\mbox{MeV}$,
which corresponds to a rather wide window of the wavelengths, $2~\mbox{fm}\lesssim \lambda_k \lesssim  
24~\mbox{fm}$. For the data in the gray shaded regions at small values of $k$, the values of the wavelengths lie 
between $\lambda_k \approx  24~\mbox{fm}$  and $\lambda_k \approx  12~\mbox{fm}$. Most likely, these 
are already unrealistically large, but we decided to presented the corresponding results for completeness. 

%%%%%%%%%%%%%%%%%%%%%%%%%%%%
%%%%%%%%%%%%%%%%%%%%%%%%%%%%

In order to obtain the numerical results in Fig.~\ref{fig:modes}, we used the lattice data for the 
transport coefficients from Ref.~\cite{Aarts:2014nba}. In this connection, it should be noted that the three selected 
choices of the temperature, $T = 200~\mbox{MeV}$, $235~\mbox{MeV}$, and $350~\mbox{MeV}$, 
correspond to $1.09 T_c$, $1.27 T_c$, and $1.9 T_c$ in the notation of Ref.~\cite{Aarts:2014nba}, 
where $T_c \approx 185~\mbox{MeV}$ is the deconfinement critical temperature obtained from
the position of the peak in the Polyakov loop susceptibility. 

As is clear from the results in Fig.~\ref{fig:modes}, all CMW-type modes are overdamped,
although not always completely diffusive. This differs somewhat from the case of the very high 
temperature and/or weak magnetic field considered in Sec.~\ref{sec:high_T}. In fact, we find that 
this is largely due to the combination of the following two effects: (i) a relatively small electrical 
conductivity of the quark-gluon plasma in the near-critical region of temperatures and (ii)  
substantial charge diffusion effects for all wave vectors allowed by the small size of the system, 
i.e., $k \gtrsim 50~\mbox{MeV}$. 

Because of a nonzero electrical conductivity, we find that one of the CMW modes becomes diffusive 
when the magnetic fields are not very strong and the wave vectors are not too large. Note that this is 
qualitatively consistent with the condition in Eq.~(\ref{condition-diffusive}) in Sec.~\ref{sec:high_T}. 
Indeed, as we see from the top row of panels in Fig.~\ref{fig:modes}, one of the modes is diffusive (i.e., 
its real part of the energy is zero) in the whole range of wave vectors shown, when the field is 
not very strong, $eB = (50~\mbox{MeV})^2$, but the temperature is high, $T = 350~\mbox{MeV}$. 
Even with decreasing the temperature, one of the modes still remains diffusive at sufficiently small 
wave vectors, namely below $k\simeq 279~\mbox{MeV}$ at $T = 235~\mbox{MeV}$ and below $k\simeq 
79~\mbox{MeV}$ at $T = 200~\mbox{MeV}$. With increasing the magnetic field, as we 
see from the second and third rows of panels in Fig.~\ref{fig:modes}, the range with one 
diffusive mode is pushed to smaller values of the wave vectors. For example, at $eB = (100~\mbox{MeV})^2$,  
the CMW is diffusive below $k\simeq 341~\mbox{MeV}$ at $T = 350~\mbox{MeV}$ and below 
$k\simeq 64~\mbox{MeV}$ at $T = 235~\mbox{MeV}$. In fact, only at the smallest value of 
temperature, $T = 200~\mbox{MeV}$, the real part of the energy is nonzero in the whole range 
of allowed wave vectors. Nevertheless, the corresponding value of the real part remains considerably 
smaller than the imaginary part. In fact, as we see from the third row of panels in Fig.~\ref{fig:modes}, 
the diffusive regime of the CMW is not completely avoided even in a rather strong magnetic field if 
the temperature stays sufficiently high. Indeed, at $eB = (200~\mbox{MeV})^2$, one of the modes 
is still diffusive below $k\simeq 73~\mbox{MeV}$ at $T = 350~\mbox{MeV}$. Only at sufficiently low 
temperatures, the CMW gradually revives and becomes a propagating mode at such an extremely 
strong field.

\begin{figure}[t]
\begin{center}
    \includegraphics[width=0.75\textwidth]{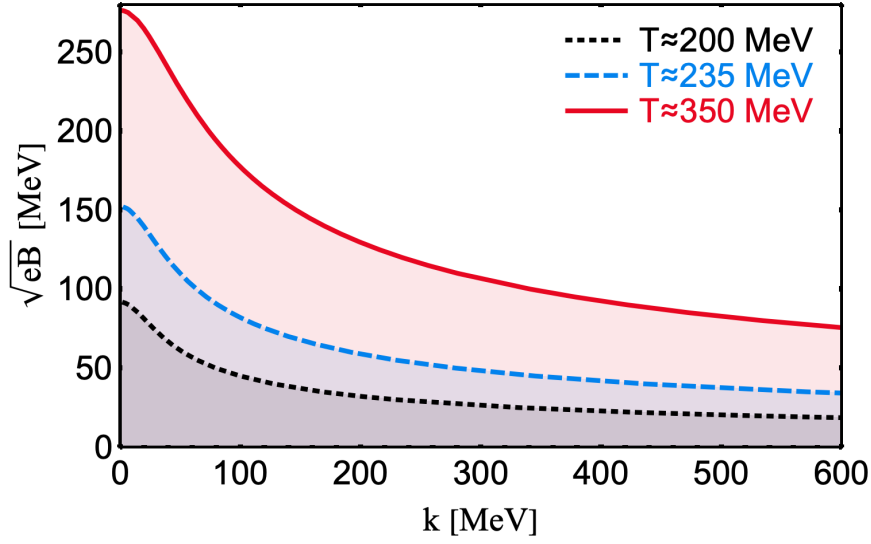}
    \caption{The graphical representation of the parameter space regions (shaded) where one of the 
    modes becomes completely diffusive. Different colors (and line types) represent the results for 
    three fixed values of temperature.}
\label{fig:phase-dig}
\end{center}
\end{figure}

The existence/absence of a completely diffusive mode in the spectrum can be easily investigated 
in the whole range of relevant model parameters. In the plane of wave vectors and magnetic field, 
the corresponding regions are presented graphically in Fig.~\ref{fig:phase-dig} for the three different 
values of temperatures. In the shaded regions (below the ``critical'' lines), the spectrum contains a diffusive 
mode. It should be pointed out that the corresponding regions agree qualitatively with the validity 
of the condition in Eq.~(\ref{condition-diffusive}).  

%%%%%%%%%%%%%%%%%%%%%%%%%%%%
%%%%%%%%%%%%%%%%%%%%%%%%%%%%

As we see from the numerical results in Fig.~\ref{fig:modes}, the collective modes are overdamped for 
all magnetic fields with the values of up to $eB \simeq (100~\mbox{MeV})^2$, i.e., even if they are not 
completely diffusive. Indeed, the ratios of the real to imaginary parts of the energies 
$E_{n}(k)/\Gamma_{n}(k)$ are less than $1$ in the whole range of the wave vectors down to the 
smallest values allowed by the system size, i.e., $k \simeq 50~\mbox{MeV}$, which corresponds 
to $\lambda_k \simeq 24~\mbox{fm}$. It easy to figure out that such strong damping cannot be 
explained by the effects of the electrical conductivity alone.

As it turns out, the charge diffusion also contributes substantially to the strong damping of the collective 
modes in a wide range of wave vectors. This is despite the fact that, in the strongly coupled quark-gluon 
plasma in the near-critical regime, the diffusion coefficient takes a rather small value, $D_f\simeq 1/(2\pi T)$, 
see Eq.~(\ref{eq-diffusion-lattice}) and Table~\ref{table:1}. By taking into account that the wave vector is 
bound from below by the inverse system size, however, one can easily see that the relevant modes are 
subject to a sizable damping.

\section{Conclusion}
\label{sec:conclusion}

In these proceedings, we critically reanalyzed the dynamics responsible for the anomalous CMW.
We found that the corresponding mode is strongly overdamped almost in all realistic regimes of 
hot plasmas after the effects of dynamical electromagnetism and charge diffusion are carefully taken into account. 

At sufficiently high temperatures and/or low magnetic fields, the propagation of the long-wavelength 
CMW is badly disrupted by the high electrical conductivity $\sigma_E$ that causes a rapid screening 
of the electric charge fluctuations. Because of such screening on the time scale $t_\textrm{scr} 
\simeq\sigma_E^{-1}$, the chiral magnetic and separation effects, which operate on the time scales 
of order $t_\textrm{CMW}\simeq 2\pi^2T^2/(3eBk)$, do not get a chance to initiate the CMW.  In 
such a regime, the corresponding mode is completely diffusive.

In the case of the nonperturbative quark-gluon plasma created in relativistic heavy-ion collisions, 
the situation is slightly more complicated because of a relatively low electrical conductivity and 
a limited range of the wave vectors allowed by the small system size. In order to study the quantitative 
properties of the anomalous collective modes in such a regime, we used the nonperturbative results for 
transport coefficients obtained in lattice calculations \cite{Aarts:2014nba}. The latter reveal that the 
electrical conductivity and charge diffusion are relatively small in the regime of near-critical temperatures
relevant for the heavy-ion experiments. Nevertheless, we still find that all CMW-type collective 
modes are strongly overdampled for the magnetic fields up to about $eB = (100~\mbox{MeV})^2$
and for the whole range of wave vectors allowed by the finite size of the system.

In connection to the quark-gluon plasma created in heavy-ion collisions, we find that even the 
relatively small electrical conductivity plays an important role and turns the long-wavelength
modes into purely diffusive ones at sufficiently low magnetic fields and/or sufficiently high 
temperatures. However, such a regime has a limited applicability for the relevant wave 
vectors allowed by the finite size of the system. By taking a rather relaxed estimate for the 
size, i.e., $R\lesssim 24~\mbox{fm}$, we found that the wave vectors are bound 
from below by $k\gtrsim 50~\mbox{MeV}$. Then, the effects of the charge diffusion 
(which grow quadratically with $k$) become very important and, in fact, often play the 
leading role in damping the collective modes. 

In the end, the combined effects of the electrical conductivity and charge diffusion cause 
a strong damping of the CMW in all realistic regimes possible in the heavy-ion collisions. 
This means that there are no theoretical foundations to expect that the CMW can exist and
produce the quadrupole charged-particle correlations \cite{Gorbar:2011ya,Burnier:2011bf}. 
By noting that a tentative detection  
of the charge-dependent flows has already been reported \cite{Ke:2012qb,Adamczyk:2013kcb,
Adamczyk:2015eqo,Adam:2015vje,Sirunyan:2017tax}, we must conclude that the corresponding 
observation is unlikely to be connected with the CMW, or any anomalous physics for that matter. 
This might explain, in fact, why the experimental effort to extract the signal from the background 
appears to be so difficult \cite{Adam:2015vje,Sirunyan:2017tax}.

In the end, it might be instructive to emphasize that here we used the hydrodynamic description 
for the collective modes. While very powerful, such an approach has some limitations. It assumes
that a local equilibrium is established in plasma. It is not applicable, therefore, for the description 
of collective modes with sufficiently short wavelengths. In the case of the CMW-type modes 
propagating along the direction of the magnetic field, however, there is a hope that the description
could be extended even down to relatively short wavelengths that are comparable to or less then the particle 
mean free path. The reason for this is rooted in the structure of the hydrodynamics equations for the CMW, which
happen to be completely decoupled from the fluid flow velocity. From a technical viewpoint, these
appear to be the same equations that come from the chiral kinetic theory and remain valid 
down to much shorter length scales.  

Last but not least, it should be emphasized that here we demonstrated that the CMW is strongly overdamped
or even diffusive in almost all regimes of hot plasmas relevant for heavy-ion physics and the early Universe.
There exists, however, one special regime in which the CMW is likely to remain a well-pronounced propagating 
mode. The corresponding regime is realized in the ultra-quantum limit with the superstrong magnetic field,
$eB\gg T^2$. In such a case, the dynamics is dominated by the lowest Landau level and all dissipative processes
are strongly suppressed, while the efficiency of the CSE and CME is maximal.

\end{document}